\documentclass[utf8]{frontiersSCNS}

\usepackage{url,lineno,microtype,subcaption}
\usepackage[onehalfspacing]{setspace}
\usepackage{adjustbox}
\usepackage{makecell}

\usepackage{bm}
\usepackage{amsmath}

\def\v{\bm{v}}
\def\u{\bm{u}}
\def\s{\bm{s}}
\def\W{\bm{W}}

\def\keyFont{\fontsize{8}{11}\helveticabold }
\def\firstAuthorLast{Shahin Heidarian {et~al.}}
\def\Authors{Shahin Heidarian\,$^{1}$,
Parnian Afshar\,$^{2}$,
Nastaran Enshaei\,$^{2}$,
Farnoosh Naderkhani\,$^{2}$,
Anastasia Oikonomou, MD\,$^{3}$,
S. Farokh Atashzar\,$^{4}$,
Faranak Babaki Fard, MD\,$^{5}$,
Kaveh Samimi, MD\,$^{6}$,
Konstantinos N. Plataniotis\,$^{7}$,
Arash Mohammadi\,$^{2,*}$,
Moezedin Javad Rafiee, MD\,$^{8}$}

\def\Address{
$^{1}$Department of Electrical and Computer Engineering, Concordia University, Montreal, QC, Canada \\
$^{2}$Concordia Institute for Information Systems Engineering (CIISE), Concordia University, Montreal, Canada\\
$^{3}$Department of Medical Imaging, Sunnybrook Health Sciences Centre, University of Toronto, Toronto, Canada \\
$^{4}$Department of Electrical, Computer Engineering \& Department of Mechanical, Aerospace Engineering, New York University, USA \\
$^{5}$Faculty of Medicine, University of Montreal, Montreal, QC, Canada\\
$^{6}$Department of Radiology, Iran university of medical science, Tehran, Iran\\
$^{7}$Department of Electrical and Computer Engineering, University of Toronto, Canada\\
$^{8}$Department of Medicine and Diagnostic Radiology, McGill University Health Center-Research Institute, Montreal, QC, Canada }
\def\corrAuthor{Arash Mohammadi}

\def\corrEmail{Concordia University, Montreal, QC,  H3G-2W1, Canada\\arash.mohammadi@concordia.ca}

\begin{document}
\onecolumn
\firstpage{1}

\title[COVID-FACT]{COVID-FACT: A Fully-Automated Capsule Network-based Framework for Identification of COVID-19 Cases from Chest CT scans}

\author[\firstAuthorLast ]{\Authors} 
\address{} 
\correspondance{} 

\extraAuth{}

\maketitle

\begin{abstract}
\section{}
The newly discovered Coronavirus Disease 2019 (COVID-19) has been globally spreading and causing hundreds of thousands of deaths around the world as of its first emergence in late 2019. The rapid outbreak of this disease has overwhelmed health care infrastructures and arises the need to allocate medical equipment and resources more efficiently. The early diagnosis of this disease will lead to the rapid separation of COVID-19 and non-COVID cases, which will be helpful for health care authorities to optimize resource allocation plans and early prevention of the disease. In this regard, a growing number of studies are investigating the capability of deep learning for early diagnosis of COVID-19. Computed tomography (CT) scans have shown distinctive features and higher sensitivity compared to other diagnostic tests, in particular the current gold standard, i.e., the Reverse Transcription Polymerase Chain Reaction (RT-PCR) test. Current deep learning-based algorithms are mainly developed based on Convolutional Neural Networks (CNNs) to identify COVID-19 pneumonia cases. CNNs, however, require extensive data augmentation and large datasets to identify detailed spatial relations between image instances. Furthermore, existing algorithms utilizing CT scans, either extend slice-level predictions to patient-level ones using a simple thresholding mechanism or rely on a sophisticated infection segmentation to identify the disease. In this paper, we propose a two-stage fully-automated CT-based framework for identification of COVID-19 positive cases referred to as the ``COVID-FACT''.  COVID-FACT utilizes Capsule Networks, as its main building blocks and is, therefore, capable of capturing spatial information.  In particular, to make the proposed COVID-FACT independent from sophisticated segmentations of the area of infection, slices demonstrating infection are detected at the first stage and the second stage is responsible for classifying patients into COVID and non-COVID cases. COVID-FACT detects slices with infection, and identifies positive COVID-19 cases using an in-house CT scan dataset, containing COVID-19, community acquired pneumonia, and normal cases. Based on our experiments, COVID-FACT achieves an accuracy of 90.82\%, a sensitivity of 94.55\%, a specificity of 86.04\%, and an Area Under the Curve (AUC) of 0.98, while depending on far less supervision and annotation, in comparison to its counterparts.

\tiny
 \keyFont{ \section{Keywords:} Capsule Networks, COVID-19, Computed Tomography Scans, Fully-Automated Classification, Deep Learning}
\end{abstract}

\section{Introduction}
The recent outbreak of the novel coronavirus infection (COVID-19) has sparked an unforeseeable global crisis since its emergence in late 2019. Resulting COVID-19 pandemic is reshaping our societies and people's lives in many ways and caused more than half a million deaths so far. In spite of the global enterprise to prevent the rapid outbreak of the disease, there are still thousands of reported cases around the world on daily bases, which raised the concern of facing a major second wave of the pandemic. Early diagnosis of COVID-19, therefore, is of paramount importance, to assist health and government authorities with developing efficient resource allocations and breaking the transmission chain.

Reverse Transcription Polymerase Chain Reaction (RT-PCR), which is currently the gold standard in diagnosing COVID-19, is time-consuming and prone to high false-negative rate~\citep{Fang2020}. Recently, chest Computed Tomography (CT) scans and Chest Radiographs (CR) of COVID-19 patients, have shown specific findings, such as bilateral and peripheral distribution of Ground Glass Opacities (GGO) mostly in the lung lower lobes, and patchy consolidations in some of the cases~\citep{Inui2020}. Diffuse distribution, vascular thickening, and fine reticular opacities are other commonly observed features of COVID-19 reported in~\citep{Bai2020,Chung2020,Shi2020,Ng2020}. Although imaging studies and their results can be obtained in a timely fashion, such features can be seen in other viral or bacterial infections or other entities such as organizing pneumonia, leading to misclassification even by experienced radiologists.

With the increasing number of people in need of COVID-19 examination, health care professionals are experiencing a heavy workload reducing their concentration to properly diagnose COVID-19 cases and confirm the results. This arises the need to distinguish normal cases and non-COVID infections from COVID-19 cases in a timely fashion to put a higher focus on COVID-19 infected cases. Using deep learning-based algorithms to classify patients into COVID and non-COVID, health care professionals can exclude non-COVID cases quickly in the first step and allow for paying more attention and allocating more medical resources to COVID-19 identified cases.
It is worth mentioning that although the RT-PCR, as a non-destructive diagnosis test, is commonly used  for COVID-19 detection, in some countries with high number of COVID-19 cases, CT imaging is widely used as the primary detection technique. Therefore, there is an unmet need to develop advanced deep learning-based solutions based on CT images to speed up the diagnosis procedure.

\subsection{Literature Review}
Convolutional Neural Networks (CNNs) have been widely used in several studies to account for the human-centered weaknesses in detecting COVID-19. CNNs are powerful models in related tasks and are capable of extracting distinguishing features from CT scans and chest radiographs~\citep{Yamashita2018}. In this regard, many studies have utilized CNNs to identify COVID-19 cases from medical images. The study by~\citep{Wang2020}, is an example of the application of CNN in COVID-19 detection, where CNN is first pre-trained on the ImageNet dataset~\citep{Krizhevsky2017}. Fine-tuning is then performed using a CR dataset. Results show an accuracy of 93.3\% in distinguishing normal, non-COVID-19 pneumonia (viral and bacterial), and COVID-19 infection cases.~\citep{Sethy2020} have also explored the same problem with the difference that the CNN is followed by a Support Vector Machine (SVM), to identify positive COVID-19 cases. Their obtained results show an overall accuracy of 95.38\%, sensitivity of 97.29\% and specificity of 93.47\%. Another study by~\citep{Mahmud2020} proposed a CNN-based model utilizing depth-wise convolutions with varying dilation rates to extract more diversified features from chest radiographs. They used a pre-trained model on a dataset of normal, viral, and bacterial pneumonia patients followed by additional fine-tuned layers on a dataset of COVID-19 and other pneumonia patients, obtaining an overall accuracy of 90.2\%, sensitivity of 89.9\%, and specificity of 89.1\%.

Chest radiograph acquisition is relatively simple with less radiation exposure than CT scans. However, a single CR image fails to incorporate details of infections in the lung and cannot provide a comprehensive view for the lung infection diagnosis. CT scan, on the other hand, is an alternative imaging modality that incorporates the detailed structure of the lung and infected areas. Unlike CR images, CT scans generate cross-sectional images (slices) to create a 3D representation of the body. Consequently, there has been a surge of interest on utilizing 2D and 3D CT images to identify COVID-19 infection. For instance,~\citep{Yang2020} proposed a DenseNet-based model to classify manually selected slices with COVID-19 manifestations and pulmonary parenchyma into COVID-19 and normal classes. The underlying study achieved an accuracy of 92\% on the patient-level classification by averaging slice-level probabilities followed by a threshold of $0.8$ on the averaged values. Furthermore, the dataset used to train and test the model does not include other types of pneumonia.  \textit{Identified Drawback 1}: Such methods require manual selecting slices demonstrating infection to feed the model, which makes the overall process time-consuming and only partially-automated. To extract features from all CT slices,~\citep{Li2020} first segmented the lung regions using a U-net based segmentation method~\citep{Ronneberger2015}, and then used them to fine-tune a ResNet50 model, which was pre-trained on natural images from the ImageNet dataset \citep{Deng2009}. Extracted features are then combined using a max-pooling operation followed by a fully connected layer to generate probability scores for each disease type. Their proposed model achieved sensitivities of 90\%, 87\%, and 94\% for COVID-19, Community Acquired Pneumonia (CAP), and non-pneumonia cases respectively. \textit{Identified Drawback 2:} Such methods combine extracted features from all slices of a patient, with or without infection, which potentially results in lower accuracy as there are numerous slices without evidence of infection in a volumetric CT scan of an infected patient.

In the study by~\citep{Hu2020}, segmented lungs are fed into a multi-scale CNN-based classification model, which utilizes intermediate CNN layers to obtain classification scores, and aggregates scores generated by intermediate layers to make the final prediction. Their proposed method achieves an overall accuracy of 87.4\% in the three-way classification.~\citep{Zhang2020} proposed a two-stage method consisting of a Deeplabv3-based lung-lesion segmentation model~\citep{Yasuda2017} followed by a 3D ResNet18 classification model \citep{Hara2017} to identify lung lesions and abnormalities and use them to classify patients into COVID-19, community acquired pneumonia, and normal findings. They manually annotated chest CT scans into seven regions to train their lung segmentation model, which is a time-consuming and sophisticated task requiring high level of thoracic radiology expertise to accomplish. Their proposed method achieves the overall accuracy of 92.49\% in both three-way and binary (COVID-19 versus others) classifications.

\subsection{Problem Statement}
At one hand, we aim to address the two identified drawbacks of the aforementioned methods. More specifically, existing solutions either require a precise annotation/labeling of lung images, which is time-consuming and error-prone, especially when we are facing a new and unknown type of disease such as COVID-19, or assign the patient-level label to all the slices. On the other hand, CNN, which is widely adopted in COVID-19 studies, suffers from an important drawback that reduces its reliability in clinical practice. CNNs are required to be trained on different variations of the same object to fully capture the spatial relations and patterns. In other words, CNNs, commonly, fail to recognize an object when it is rotated or transformed. In practice, extensive data augmentation and/or adoption of huge data resources are needed to compensate for the lack of spatial interpretation. As COVID-19 is a relatively new phenomenon, large datasets are not easily accessible, especially due to strict privacy preserving constraints. Furthermore, most COVID-19 cases have been reported with a specific infection distribution in their image~\citep{Bai2020,Chung2020,Shi2020,Ng2020}, which makes capturing spatial relations in the image highly important.

\subsection{Contributions}
As stated previously, structure of infection spread in the lung for COVID-19 is not yet fully understood given its recent and abrupt emergence. Furthermore, COVID-19 has a particular structure in affecting the lung, therefore, picking up those spatial structures are significantly important. Capsule Networks (CapsNets)~\citep{Hinton2018}, in contrast to CNNs, are equipped with routing by agreement process enabling them to capture such spatial patterns.  Even without a large dataset, capsules interpret the object instantiation parameters, besides its existence, and by reaching a mutual agreement, higher-level objects are developed from lower-level ones. The superiority of Capsule Networks over their counterparts has been shown in different medial image processing problems~\citep{Afshar2018,Afshar2019,Afshar2019a,Afshar2020,Afshar2020b,Afshar2020c}. Recently, we proposed a Capsule Network-based framework~\citep{Afshar2020a}, referred to as the COVID-CAPS, to identify COVID-19 cases from chest radiographs, which achieved an accuracy of 98.3\%, a specificity of 98.6\%, and a sensitivity of 80\%. As stated previously, CT imaging is superior for COVID-19 detection and diagnosis purposes when compared to chest radiographs. However, as in the case of CT imaging, we are dealing with 3D inputs and several slices per patient (compared to one chest radiograph per patient), the learning process is significantly more challenging. As such, accuracies of deep models trained over CT scans cannot be directly compared with those obtained based on chest radiographs.

Following our previous study on chest radiographs, in the present study, we take one step forward and propose a fully automated two-stage Capsule Network-based framework, referred to as the COVID-FACT, to identify COVID-19 patients using chest CT images. Based on our in-house dataset, COVID-FACT achieves an accuracy of 90.82\%, sensitivity of 94.55\%, specificity of 86.04\%, and Area Under the Curve (AUC) of 0.98. We developed two variants of the COVID-FACT, one of which is fed with the whole chest CT image, while the other one utilizes the segmented lung area as the input. In the latter case, instead of using an original chest CT image, first a segmentation model~\citep{Hofmanninger2020} is applied to extract the lung region, which is then provided as input to the COVID-FACT. This will be further clarified in Section~\ref{detail-method}. Experimental results show that the model coupled with lung segmentation achieves the same overall accuracy compared to the other COVID-FACT variation working with original images. However, using the segmented lung regions increases the sensitivity and AUC from 92.72\% and 0.95 to 94.55\% and 0.98, respectively, while slightly decreasing the specificity from 88.37\% to 86.04\%. The $95\%$ Confidence Interval (CI) is also provided for all performance matrics using the methods described in~\citep{Brown2001,Hanley1982}.

COVID-FACT benefits from a two-stage design, which is of paramount importance in COVID-19 detection using CT scans, as a CT examination is typically associated with hundreds of slices that cannot be analyzed at once. At the first stage, the proposed COVID-FACT detects slices demonstrating infection in a 3D volumetric CT scan to be analyzed and classified at the next  stage. At the second stage, candidate slices detected at the previous stage are classified into COVID and non-COVID (community acquired pneumonia and normal) cases and a voting mechanism is applied to generate the classification scores in the patient level. COVID-FACT's two-stage architecture has the advantage of being trained by even weakly labeled dataset, as errors at the first stage can be compensated at the second stage. As a result, COVID-FACT does not require any infection annotation or a very precise slice labeling, which is a valuable asset due to the limited knowledge and experience on the novel COVID-19 disease.
The trained COVID-FACT model and the code to implement the corresponding fully automated framework are available publicly for open access at https://github.com/ShahinSHH/COVID-FACT.

The reminder of the paper is organized as follows: Section~\ref{materials} describes the dataset and imaging protocol used in this study. Section~\ref{detail-method} presents a brief description of Capsule Networks and explains the proposed COVID-FACT in details. Experimental results and model evaluation are presented in Section~\ref{results}. Finally, Section~\ref{discussion} concludes the work.

\section{Materials and Equipment} \label{materials}
In this section, we will explain the in-house dataset used in this study, along with the associated imaging protocol.

\subsection{Dataset}
\begin{figure*}
\centering
\caption{ \textbf{A}, \textbf{B}: Infected and non-infected sample slices in a COVID-19 case; \textbf{C}, \textbf{D}: Infected and non-infected sample slices in a non-COVID Pneumonia case. }
\includegraphics[scale=.5]{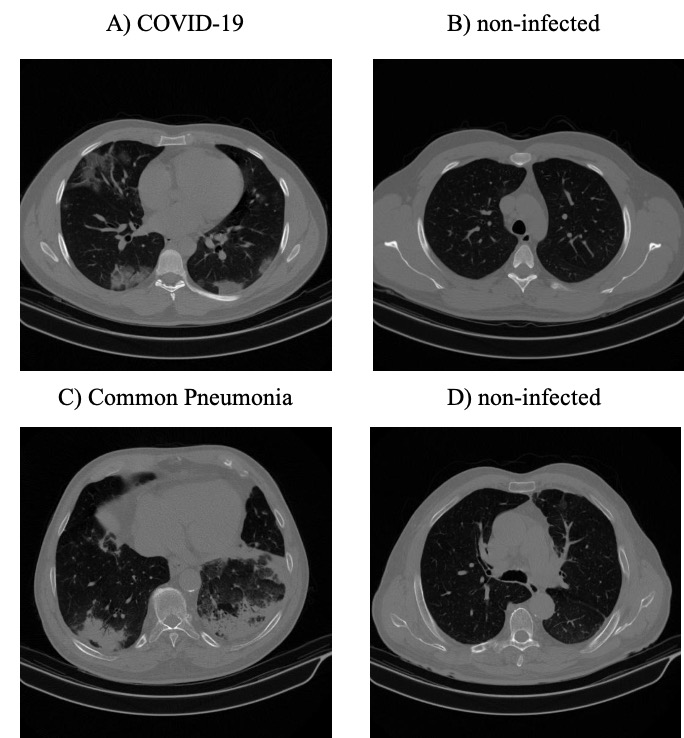}
\label{fig:sample}
\vspace{-.1in}
\end{figure*}

This research work is performed based on the policy certification number $30013394$ of Ethical acceptability for secondary use of medical data approved by Concordia University.
The dataset used in this study, referred to as the ``COVID-CT-MD"~\citep{Afshar2020d}, contains volumetric chest CT scans of 171 patients positive for COVID-19 infection, 60 patients with community acquired pneumonia, and 76 normal patients acquired from April 2018 to May 2020. The average age of patients is $50\pm16$ including 183 men and 124 women.
Diagnosis of COVID-19 infection is based on positive real-time reverse transcription polymerase chain reaction (rRT-PCR) test results, clinical parameters, and CT scan manifestations by a thoracic radiologist, with 20 years of experience in thoracic imaging. Community acquired pneumonia and normal cases were included from another study and the diagnosis was confirmed using clinical parameters, and CT scans. A subset of 55 COVID-19, and 25 community acquired pneumonia cases were analyzed  by the radiologist to identify and label slices with evidence of infection as shown in Figure~\ref{fig:sample}. This labeling process focuses more on distinctive manifestations rather than slices with minimal findings. The labeled subset of the data contains 4,962 number of slices demonstrating infection and 18,447 number of slices without infection. The data is then used to train and validate the first stage of our proposed COVID-FACT model for extracting slices demonstrating infection from a volumetric CT data to be used in the second classification stage. We divided this subset into three separate components for training, validation, and testing. 60\% of the labeled data is used for training, 10\% for validation, and 30\% for the test. The unlabeled subset is divided with the same proportion and used along with the labeled data to develop the second stage of the COVID-FACT model and evaluate the overall method.

\subsection{Imaging Protocol}
All CT examinations have been acquired using a single CT scanner with the same acquisition setting and technical parameters, which are presented in Table~\ref{tab:ct}, where kVP (kiloVoltage Peak) and Exposure Time affect the radiation exposure dose, while Slice Thickness and Reconstruction Matrix represent the axial resolution and output size of the images,  respectively~\cite{Raman2013}. Next, we describe the proposed  COVID-FACT framework  followed by the experimental results.

\begin{table*}[t!]
\centering
\caption{Imaging device and settings used to acquire the in-house dataset.}
\label{tab:ct}
\vspace{.1in}
\renewcommand{\arraystretch}{2}
\begin{adjustbox}{width=1\textwidth}
\begin{tabular}{|c|c|c|c|c|c|c|c|}
\hline
\textbf{Scanner Manufacturer and Model} & \textbf{Slice Thickness (mm)} & \textbf{Image Type} & \textbf{kVP (kV)} & \textbf{Exposure Time (ms)} & \textbf{Reconstruction Matrix} & \textbf{Window Center} & \textbf{Window Width} \\[1ex]
\hline
SIEMENS,
SOMATOM Scope & $2$ & Axial & $110$ & $600$ & $512\times512$ & $[50, -600]$ & $[350, 1200]$ \\[1ex]
\hline
\end{tabular}
\end{adjustbox}
\end{table*}

\section{Method}\label{detail-method}
The COVID-FACT framework is developed to automatically distinguish COVID-19 cases from other types of pneumonia and normal cases using volumetric chest CT scans. It utilizes a lung segmentation model at a pre-processing step to segment lung regions and pass them as the input to the two-stage Capsule Network-based classifier. The first stage of the COVID-FACT extracts slices demonstrating infection in a CT scan, while the second stage uses the detected slices in first stage to classify patients into COVID-19 and non-COVID cases. Finally, the Gradient-weighted Class Activation Mapping (Grad-CAM) localization approach~\citep{Selvaraju2017} is incorporated into the model to highlight important components of a chest CT scan, that contribute the most to the final decision.

In this section, different components of the proposed COVID-FACT are explained. First, Capsule Network, which is the main building block of our proposed approach, is briefly introduced. Then the lung segmentation method is described, followed by the details related to the first and second stages of the COVID-FACT architecture. Finally, the Grad-CAM localization mapping approach is presented.

\subsection{Capsule Networks}
A Capsule Network (CapsNet) is an alternative architecture for CNNs with the advantage of capturing hierarchical and spatial relations between image instances. Each Capsule layer utilizes several capsules to determine existence probability and pose of image instances using an instantiation vector. The length of the vector represents the existence probability and the orientation determines the pose. Each Capsule $i$ consists of a set of neurons, which collectively create the instantiation vector $\u_i $ for the associated instance. Capsules in lower layers try to predict the output of Capsules in higher levels using a trainable weight matrix $\W_{ij}$ as follows
\begin{equation}
\hat{\u}_{j|i} = \W_{ij}\u_i,
\end{equation}
where $\hat{\u}_{j|i}$ is the predicted output of Capsule $j$ in the next layer by the Capsule $i$ in the lower layer. The association between the prediction $\hat{\u}_{j|i}$ and the actual output of Capsule $j$, denoted by $\v_j$, is determined by taking the inner product of $\hat{\u}_{j|i}$ and $\v_j$. The higher the inner product, the more contribution of the lower level capsules to the higher level one. The contribution of Capsule $i$ to the output of the Capsule $j$ in the next  layer is determined by a coupling coefficient $c_{ij}$, trained over a course of few iterations known as the ``Routing by Agreement"  given by
\begin{eqnarray}
a_{ij} &=& \v_j.\hat{\u}_{j|i},\\
b_{ij} &=& b_{ij} + a_{ij},\\
c_{ij} &=& \frac{\exp(b_{ij})}{\sum_k \exp(b_{ik})},\\
\s_j &=& \sum_i{c_{ij}\hat{\u}_{j|i}},\\
\text{and} \qquad \v_j &=& \frac{\left \| \s_j \right \|^{2}}{1+\left \| \s_j \right \|^{2}}\frac{\s_j}{\left \| \s_j \right \|}, \label{squash}
\end{eqnarray}
where $a_{ij}$ is referred to as the agreement coefficient between the prediction and actual output, and $b_{ij}$ denotes the log prior of the coupling coefficient $c_{ij}$. Vector $\s_j$ denotes the Capsule output before applying the squashing function. As the length of output vectors represents probabilities, the ultimate output of Capsule $j$ ($\v_j$) is obtained by squashing $\s_j$ between 0 and 1 using the squashing function defined in Eq.~\eqref{squash}.
In order to update weight matrix $\W_{ij}$ through a backward training process, the loss function is calculated for each Capsule $k$ as follows
\begin{equation}
l_{k}=T_{k}\max(0,m^+-||\v_k||)^2+\lambda(1-T_k)\max(0,||\v_k||-m^-)^2,
\end{equation}
where $T_k$ is 1 when the class $k$ is present and 0 otherwise. $m^+$, $m^-$, and $\lambda$ are hyper parameters of the model and are originally set to 0.9, 0.1, and 0.5,  respectively. The overall loss is the summation of all losses calculated for all Capsules.


\subsection{Proposed COVID-FACT}
\begin{figure*}
\centering
\caption{{The two-stage architecture of the proposed COVID-FACT.}}
\includegraphics[width=1\textwidth]{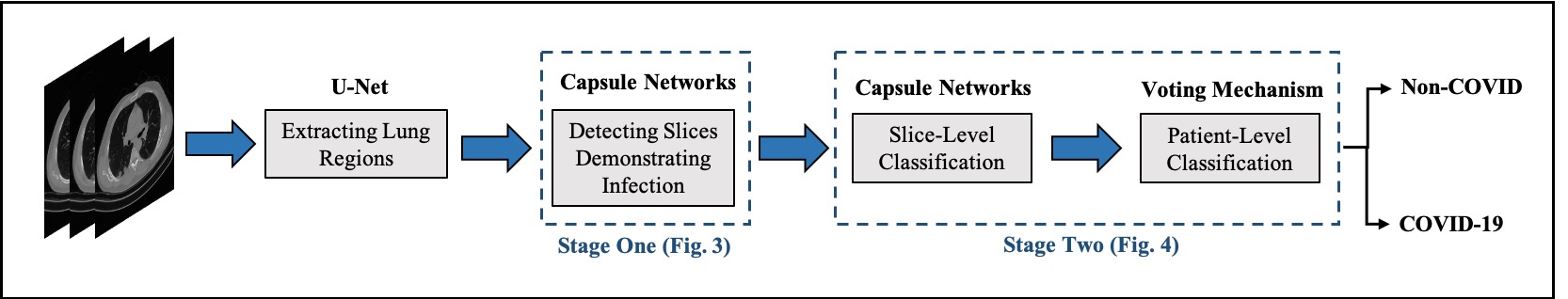}
\label{fig:method}
\vspace{-.1in}
\end{figure*}

The overall architecture of the COVID-FACT is illustrated in Figure~\ref{fig:method}, which consists of a lung segmentation model at the beginning followed by two Capsule Network-based models and an average voting mechanism coupled with a thresholding approach to generate patient-level classification results. The three components of the COVID-FACT are as follows:
\begin{itemize}
\item{\textbf{Lung Segmentation:}} The input of the COVID-FACT is the segmented lung regions identified by a U-net based segmentation model~\citep{Hofmanninger2020}, referred to as the ``U-net (R231CovidWeb)'', which has been initially trained on a large and diverse dataset including multiple pulmonary diseases, and fine-tuned on a small dataset of COVID-19 images. The Input of the U-net (R231CovidWeb) model is a single slice with the size of $512\times 512$. The output is the lung tissues, which will be normalized between 0 and 1 to generalize the features and help the model to perform more effectively. Following the literature~\citep{Zhang2020,Hu2020}, we down-sampled the output from $512 \times 512$ to $256 \times 256$ size to reduce the complexity and memory allocation without losing any significant information. Slices with no detected lung regions are removed and the remaining are fed into the first stage of the COVID-FACT model.
\item{\textbf{COVID-FACT's Stage One:}} The first stage of the COVID-FACT, shown in Figure~\ref{fig:stage1} is responsible to identify slices demonstrating infection (by COVID-19 or other types of pneumonia). Using this stage, we discard slices without infection and focus only on the ones with infection. Intuitively speaking, this process is similar in nature to the way that radiologists analyze a CT scan. When radiologists review a CT scan containing numerous consecutive cross-sectional slices of the body, they identify the slices with an abnormality in the first step, and analyze the abnormal ones to diagnose the disease in the next step. Existing CT-based deep learning processing methods either use all slices as a 3D input to a classifier, or classify individual slices and transform slice-level predictions to the patient-level ones using a threshold on the entire slices~\citep{Rahimzadeh2020}. Determining a threshold on the number or percentage of slices demonstrating infection over the entire slices is not precise, as most pulmonary infections have different stages with involvement of different lung regions \citep{Yu2020}. Furthermore, a CT scan may contain different number of slices depending on the acquisition setting, which makes it impossible to find such a threshold. In most methods passing all slices as a 3D input to the model, the input size is fixed and the model is trained to assign higher scores to slices demonstrating infection. However, the performance of such models will be reduced when testing on a dataset other than the dataset on which they are originally trained~\citep{Zhang2020}.

The model used in stage one of the proposed COVID-FACT is adapted from the COVID-CAPS model presented in our previous work \citep{Afshar2020a}, which was developed to identify COVID cases from chest radiographs. The first stage consists of four convolutional layers and three capsule layers. The first and second layers are convolutional ones followed by a batch-normalization. Similarly, the third and fourth layers are convolutional ones followed by a max-pooling layer. The fourth layer, referred to as the primary Capsule layer, is reshaped to form the desired primary capsules. Afterwards, three capsule layers perform sequential routing processes. Finally, the last Capsule layer represents two classes of infected and non-infected slices. The input of  stage one is set of CT slices corresponding to a patient, and the output is slices of the volumetric CT scan demonstrating infection. The output of stage one may vary in size for each patient due to different areas of lung involvement and phase of infection.

In order to cope with our imbalanced training dataset, we modified the loss function, so that a higher penalty rate is given to the false positive (infected slices) cases. The loss function is modified as follows
\begin{equation}
\label{loss-function}
loss = \frac{N^{+}}{N^{+}+N^{-}}\times loss^{-} + \frac{N^{-}}{N^{+}+N^{-}}\times loss^{+},
\end{equation}
where $N^{+}$ is the number of positive samples, $N^{-}$ is the number of negative samples, $loss^{+} $ denotes the loss associated with positive samples, and $loss^{-}$  denotes the loss associated with negative samples.
%
\begin{figure*}
\centering
\caption{ Architecture of the COVID-FACT at stage one.}
\includegraphics[width=1\textwidth]{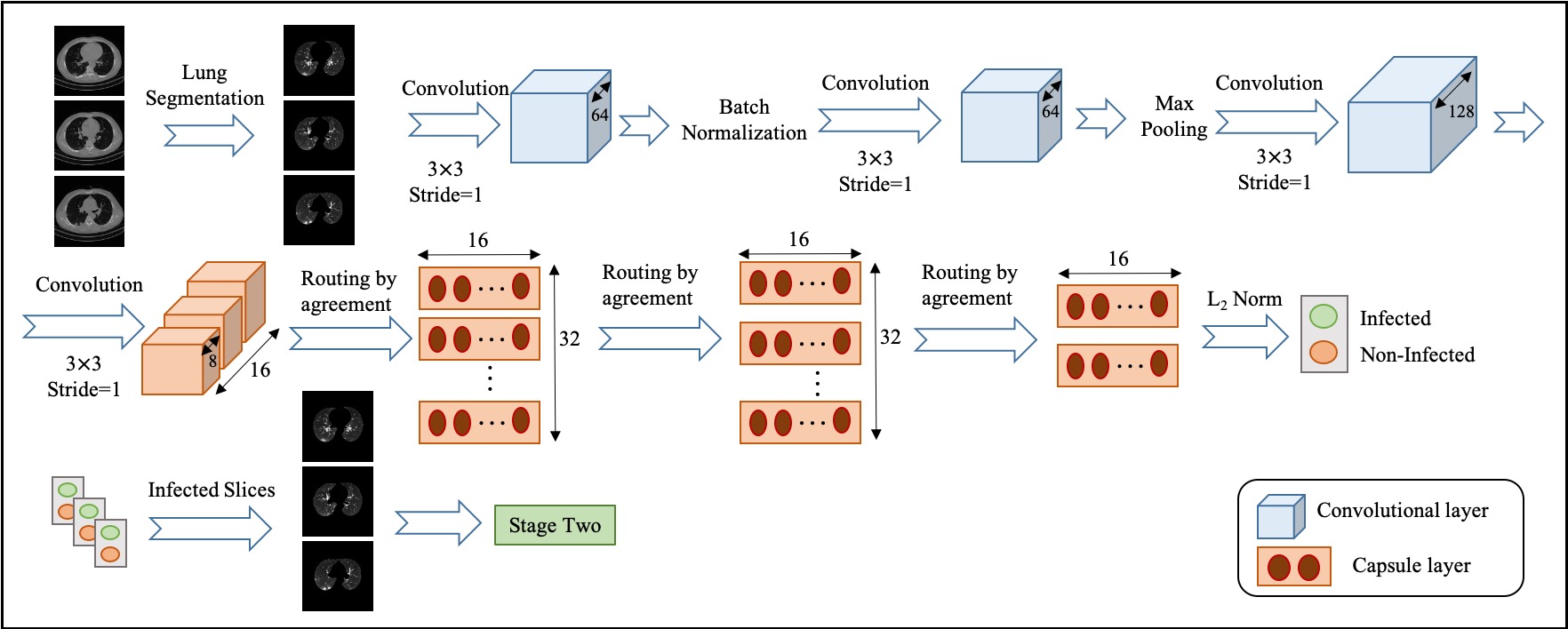}
\label{fig:stage1}
\vspace{-.1in}
\end{figure*}
%
\item{\textbf{COVID-FACT's Stage Two:}} As mentioned earlier, we need to apply classification methods on a subset of slices demonstrating infection rather than on the entire slices in a CT scan. It is worth noting that, lung segmentation (i.e., extracting lung tissues) is performed in one of the variants of the COVID-FACT as a pre-processing step. The first stage of the COVID-FACT, on the other hand, is tasked with this specific issue of extracting slices demonstrating infections.

The second stage of the COVID-FACT takes candidate slices of a patient detected in stage one as the input, and classifies them into one of COVID-19 or non-COVID (including normal and pneumonia) classes, i.e., we consider a binary classification problem. Stage two is a stack of four convolutional and two capsule layers shown in Figure~\ref{fig:stage2}. The output of the last capsule indicates classification probabilities in the slice-level. An average voting function is applied to the classification probabilities, in order to aggregate slice-level values and find the patient-level predictions as follows
\begin{equation}
\label{eqn:vote}
P(p_{k}\in c) =\frac{1}{L_{k}} \sum_{i=1}^{L_{k}}P(s_{i}^{k}\in c),
\end{equation}
where $P(p_{k}\in c)$ refers to the probability that patient $k$ belongs to the target class $c$ (e.g., COVID-19), $L_{k}$ is the total number of slices detected in stage one for patient $k$, and $P(s_{i}^{k}\in c)$ refers to the probability that the $i^{\text{th}}$ slice detected for patient $k$ belongs to the target class $c$. It is worth noting that while, initially, the COVID-FACT performs slice-level classification in its second stage, the output is patient-level classification (through its voting mechanism), which is on par with other works that COVID-FACT is compared with. As a final note to our discussion, we would like to add that, corona virus infection is, typically, distributed across the lung volume as such manifests itself in several CT slices. Therefore, having a single slice identified as COVID-19 infection can not necessarily lead to a positive COVID-19 detection.

Similar to stage one, the loss function modification in Eq.~\eqref{loss-function} is used in the training phase of Stage two. The default cut-off probability of 0.5 is chosen in Stage two to distinguish COVID-19 and non-COVID cases. However, it is worth mentioning that the main concern in the clinical practice is to have a high sensitivity in identifying COVID-19 positive patients, even if the specificity is not very high. As such, the classification cut-off probability can be modified by physicians using the ROC curve shown in Figure~\ref{fig:roc} in order to provide a desired balance between the sensitivity and the specificity (e.g., having a high sensitivity while the specificity is also satisfying). In other words, physicians can decide how much certainty is required to consider a CT scan as a COVID-19 positive case. By choosing a cut-off value higher than 0.5, we can exclude those community acquired pneumonia cases that contain highly overlapped features with COVID-19 cases. On the other hand, by selecting a lower cut-off value, we will allow more cases to be identified as a COVID-19 case.

To further improve the ability of the proposed COVID-FACT model to distinguish COVID-19 and non-COVID cases and attenuate effects of errors in the first stage, we classify all patients with less than 3\% of slices demonstrating infection in the entire volume as a non-COVID case. These cases are more likely normal cases without any slices with infection. The few slices with infection identified for these cases might be due to the model error in the first stage, non-infectious abnormalities such as pulmonary fibrosis, or motion artifacts in the original images, which will be covered by this threshold. Based on \citep{Yu2020}, it can be interpreted that 4\% lung involvement is the minimum percentage for COVID-19 positive cases. In addition, the minimum percentage of slices demonstrating infection detected by the radiologist in our dataset is 7\%, and therefore 3\% would be a safe threshold to prevent mis-classifying infected cases as normal.

\item{\textbf{Grad-CAM:}} Using the Grad-CAM approach, we can visually verify the relation between the model's prediction and the features extracted by the intermediate convolutional layers, which ultimately leads to a higher level of interpretability of the model. Grad-CAM's outcome is a weighted average of the feature maps of a convolutional layer, followed by a Rectified Linear Unit (ReLU) activation function,~i.e.,
\begin{equation}
\label{eqn:gradcam}
L_{Grad-CAM}^{c} = RelU\left ( \sum_{k}\alpha _{k}^{c}A^{k} \right),
\end{equation}
where $L_{Grad-CAM}^{c}$ refers to the Grad-CAM's output for the target class $c$; $\alpha _{k}^{c}$ is the importance weight for the feature map $k$ and the target class $c$, and;  $A^{k}$ refers to the feature map $k$ of a convolutional layer. The weights $\alpha _{k}^{c}$ are obtained based on the gradients of the probability score of the target class with respect to an intermediate convolutional layer followed by a global average pooling function  as follows
\begin{equation}
\label{eqn:gradweight}
\alpha _{k}^{c} = \frac{1}{Z}\sum_{i}^{}\sum_{j}^{}\frac{\partial y^{c} }{\partial A_{ij}^{k}},
\end{equation}
where $y^{c}$ is the prediction value (probability) for target class $c$, and $Z$ refers to the total number of feature maps in the convolutional layer.

\begin{figure*}
\caption{ Architecture of the COVID-FACT at  stage two. }
\centering
\includegraphics[width=1\textwidth]{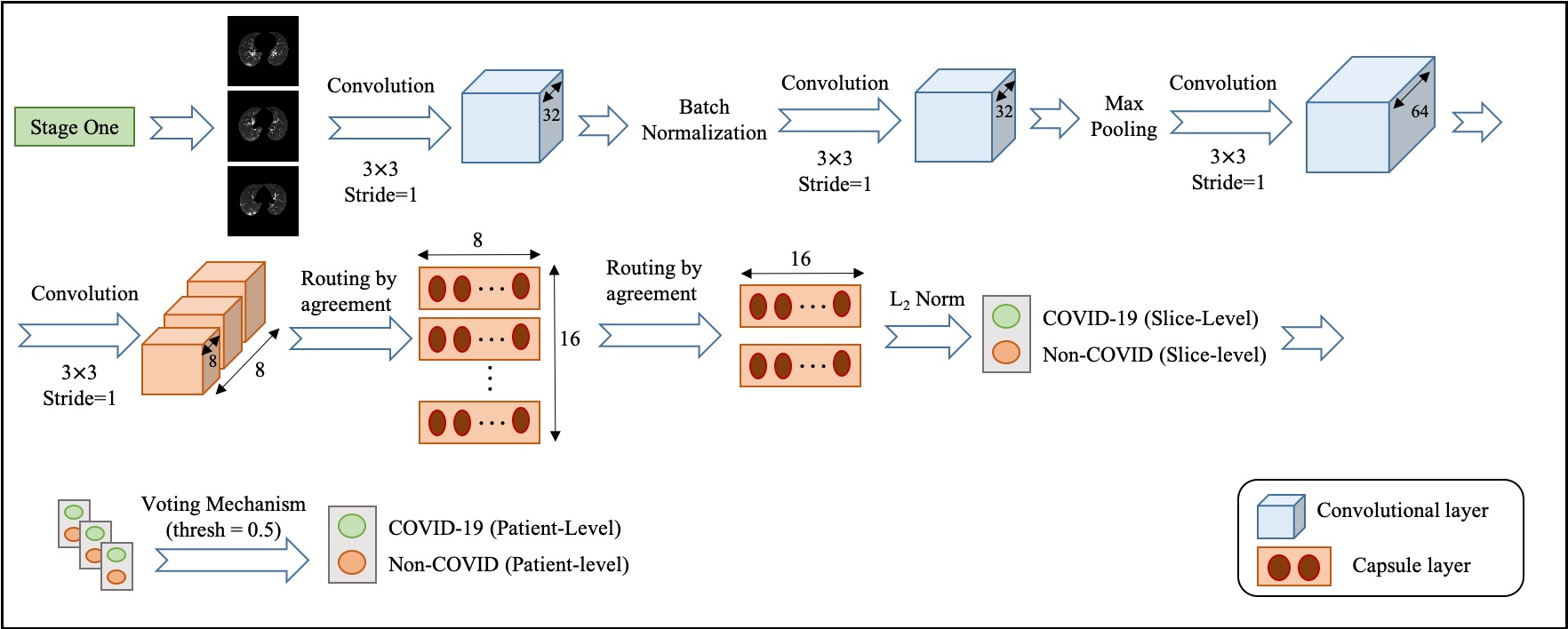}
\label{fig:stage2}
\vspace{-.1in}
\end{figure*}
\end{itemize}

\section{Experimental Results} \label{results}

\begin{figure*}
\caption{ROC curve of the proposed COVID-FACT. Values in brackets show $95\%$ confidence interval.}
\centering
\includegraphics[scale=.35]{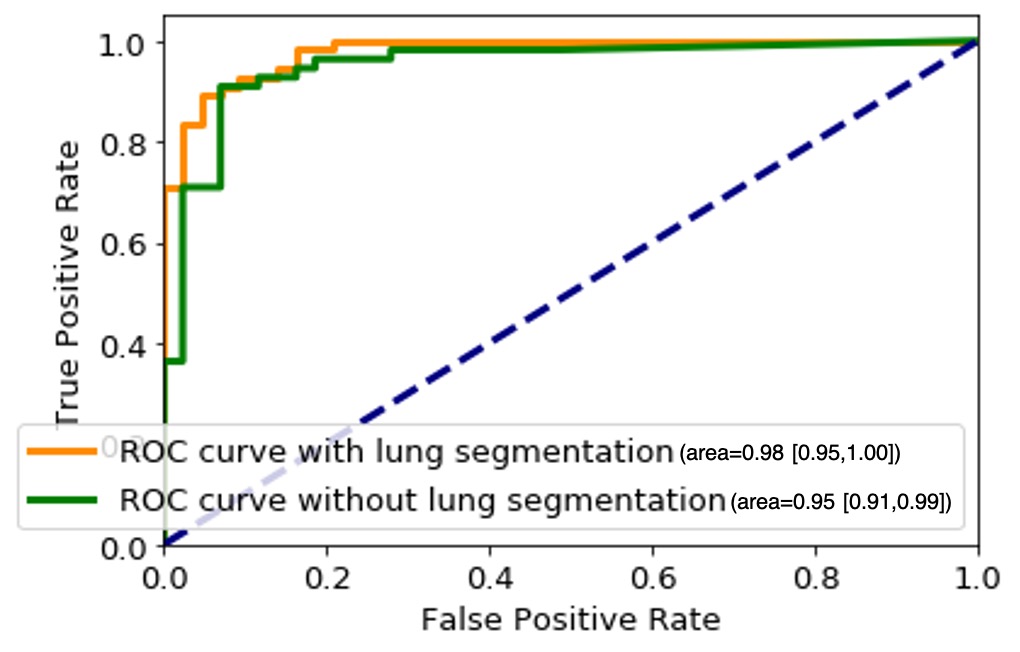}
\label{fig:roc}
\vspace{-.1in}
\end{figure*}

The proposed COVID-FACT is tested on the in-house dataset described earlier in Section~\ref{materials}. The testing set contains 53 COVID-19 and 43 non-COVID cases (including 19 community acquired pneumonia and 24 normal cases). We used the Adam optimizer with the initial learning rate of $1e-4$, batch size of 16, and 100 epochs. The model with the minimum loss value on the validation set was selected to evaluate the performance of the model on the test set.  The proposed COVID-FACT method achieved an accuracy of 90.82\%, sensitivity of 94.55\%, specificity of 86.04\%, and AUC of 0.97. The obtained ROC curve is shown in Figure~\ref{fig:roc}.

In a second experiment, we trained our model using the complete CT images without segmenting the lung regions. The obtained model reached an accuracy of 90.82\%, sensitivity of 92.72\%, specificity of 88.37\%, and AUC of 0.95. The corresponding ROC curve is shown in Figure~\ref{fig:roc}. This experiment indicates that segmenting lung regions in the first step will increase the sensitivity and AUC from 92.72\% and 0.95 to 90.82\% and 0.98 respectively, while slightly decreases the specificity from 88.37\% to 86.04\%. Although the numerical results show a slight improvement achieved by segmenting the lung regions, further investigating the sources of errors demonstrates the superiority of using segmented lung regions over the original CT scans. In the COVID-FACT model using lung segmented regions, none of COVID-19 and community acquired pneumonia cases have been mis-classified as a normal case by the 3\% thresholding after the first stage, and 95.84\% (23/24) of normal cases have been identified correctly using this threshold, while for the model without the lung segmentation, there is one mis-classification of a COVID-19 case by the 3\% thresholding, and 91.66\% (22/24) of normal cases were identified correctly using this threshold.

Furthermore, we compared  performance of the Capsule Network-based framework of COVID-FACT with a CNN-based alternative to demonstrate the effectiveness of Capsule Networks and their superiority over CNN in terms of number of trainable parameters and accuracy. In other words, the CNN-based alternative model has the same front-end (convolutional layers) as that of COVID-FACT in both stages. However, the Capsule layers are replaced by fully connected layers including 128 neurons for intermediate layers and 2 neurons for the last layer at each stage. The last fully connected layer in each stage is followed by a sigmoid activation function and the remaining modifications and hyper-parameters are kept the same as used in COVID-FACT. The CNN-based COVID-FACT achieved an accuracy of 71.43\%, sensitivity of 81.82\%, and specificity of 58.14\%. The COVID-FACT performance, and number of trainable parameters for examined models are presented in Table~\ref{tab:performance}.

As mentioned earlier, the ROC curve provides physicians  with a precious tool to modify the sensitivity/specificity balance based on their preference by changing the classification cut-off probability. To elaborate this point, we changed the default cut-off probability from $0.5$ to $0.75$ and reached an accuracy of 91.83\%, a sensitivity of 90.91\%, and a specificity of 93.02\%. Further increasing the cut-off probability to $0.8$ results in the same accuracy of 91.83\%, a lower sensitivity of 89.01\%, and a higher specificity of 95.34\%. The performance of the COVID-FACT for different values of cut-off probability are presented in Table~\ref{tab:cutoff}. While the performance of the COVID-FACT is evaluated by its final decision made in the second stage, the first stage plays a crucial role in the overall accuracy of the model. As such, the performance of the COVID-FACT in the first stage is also reported. The COVID-FACT achieves an accuracy of 93.14\%, a sensitivity of 90.75\%, a specificity of 94.01\%, and an AUC of 0.96 in detecting slices demonstrating infection in a volumetric CT scan.

\begin{table*}[t!]
\centering
\caption{Performance of COVID-FACT for different values of cut-off probability. Values in parenthesis show $95\%$ confidence interval.}
\label{tab:cutoff}
\vspace{.1in}
\renewcommand{\arraystretch}{2}

\begin{tabular}{|c|c|c|c|c|c|c|}
\hline
\textbf{Cut-off Probability} & \textbf{0.5} & \textbf{0.6} & \textbf{0.7} &\textbf{0.75} &\textbf{0.8}  \\
\hline
\textbf{Accuracy (\%)}  & \makecell{$90.82$\\$(90.24,91.40)$} & \makecell{\textbf{91.83}\\$(91.28,92.38)$} & \makecell{$90.82$\\$(90.24,91.40)$} & \makecell{\textbf{91.83}\\$(91.28,92.38)$} & \makecell{\textbf{91.83}\\$(91.28,92.38)$} \\
\hline
\textbf{Sensitivity (\%)} & \makecell{\textbf{94.55}\\$(93.74,95.36)$} & \makecell{$92.73$\\$(91.80,93.66)$} & \makecell{$90.91$\\$(89.89,91.93)$} & \makecell{$90.91$\\$(89.89,91.93)$} & \makecell{$89.01$\\$(87.90,90.12)$}  \\
\hline
\textbf{Specificity (\%)}& \makecell{$86.04$\\$(84.46,8762)$} & \makecell{$90.70$\\$(89.38,92.02)$} & \makecell{$90.70$\\$(89.38,92.02)$} & \makecell{$93.02$\\$(91.86,94.18)$} & \makecell{\textbf{95.34}\\$(94.38,96.30)$}  \\
\hline
\end{tabular}

\end{table*}

\begin{figure*}
\caption{Localization heatmaps for the second and forth convolutional layers of the first stage obtained by the Grad-CAM for two slices. }
\centering
\includegraphics[scale=.50]{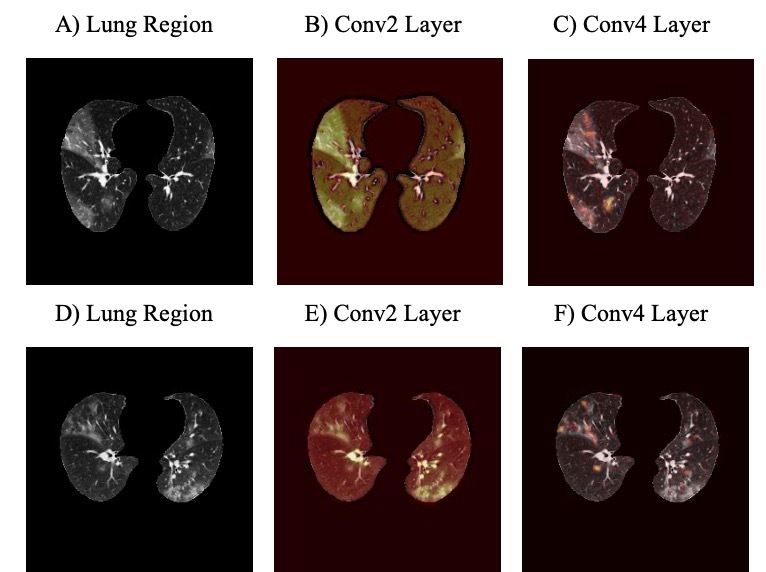}
\label{fig: heatmap}
\vspace{-.1in}
\end{figure*}

The localization maps generated by the Grad-CAM method are illustrated in Figure~\ref{fig: heatmap} for the second and fourth convolutional layers in the first stage of the COVID-FACT. It is evident in Figure~\ref{fig: heatmap} that the COVID-FACT model is looking at the right infectious areas of the lung to make the final decision. Due to the inherent structure of the Capsule layers, which represent image instances separately, their outputs cannot be superimposed over the input image. Consequently, in this study, the Grad-CAM localization maps are presented only for convolutional layers.

\begin{table*}[t!]
\centering
\caption{Results obtained from COVID-FACT and the alternative CNN-based model. Values in parenthesis show $95\%$ confidence interval.}
\label{tab:performance}
\vspace{.1in}
\renewcommand{\arraystretch}{2}
\begin{adjustbox}{width=1\textwidth}
\begin{tabular}{|c|c|c|c|c|c|}
\hline
\textbf{Method} & \textbf{Accuracy(\%)} & \textbf{Sensitivity(\%)} & \textbf{Specificity(\%)} & \textbf{AUC} & \textbf{Trainable Parameters} \\[1ex]
\hline
\textbf{COVID-FACT with Lung Segmentation} & \makecell{\textbf{90.82}\\$(90.24,91.40)$} &\makecell{\textbf{94.55}\\$(93.74,95.36)$} &  \makecell{$86.04$\\$(84.46,8762)$} & \makecell{\textbf{0.98}\\$(0.95,1.00)$} &\textbf{406,880} \\[1ex]
\hline
\textbf{COVID-FACT without Lung Segmentation}&  \makecell{\textbf{90.82}\\$(90.24,91.40)$} & \makecell{$92.72$\\$(91.79,93.65)$} & \makecell{\textbf{88.37}\\$(86.91,89.83)$} & \makecell{$0.95$\\$(0.91,0.99)$} & \textbf{406,880} \\
\hline
\textbf{CNN-based COVID-FACT} & \makecell{$71.43$\\$(70.53,72.33)$} & \makecell{$81.82$\\$(80.45,83.19)$} & \makecell{$58.14$\\$(55.89,60.39)$} & \makecell{$0.67$\\$(0.56,0.78)$} & $365,806,660$ \\
\hline
\end{tabular}
\end{adjustbox}
\end{table*}

\section{Discussion}
\label{discussion}
\indent

In this study, we proposed a fully automated Capsule Network-based framework, referred to as the COVID-FACT, to diagnose COVID-19 disease based on chest CT scans. The proposed framework consists of two stages, each of which containing several layers of convolutional and Capsule layers. COVID-FACT is augmented with a thresholding method to classify CT scans with zero or very few slices demonstrating infection as non-COVID patients, and an average voting mechanism coupled with a  thresholding approach is embedded to extend slice-level classification into patient-level ones. Experimental results indicate that the COVID-FACT achieves a satisfactory performance, in particular a high sensitivity with far less trainable parameters, supervision requirements, and annotations compared to its counterparts.

We further investigated sources of errors to determine the limitations and possible improvements. 26.31\% (5/19) of community acquired pneumonia cases have been mis-classified as a COVID-19 case, while only 4.16\% (1/24) of normal cases have been mis-classified. As mentioned earlier, none of COVID-19 and community acquired pneumonia cases have been mis-classified as a normal case by the 3\% thresholding after the first stage, and 95.84\% (23/24) of normal cases have been identified correctly using such an approach. This indicates the capability of COVID-FACT to identify normal cases in the first stage, which is very helpful for physicians and radiologists to exclude normal cases at the very beginning of their study. We also identified that errors in stage one are mainly caused by non-infectious abnormalities such as pulmonary fibrosis and motion artifacts. Errors in stage two are mostly caused by mis-classification of community acquired pneumonia cases as COVID-19 cases due to the significant overlap between these two types of infection. It is worth mentioning that during the labeling process accomplished by the radiologist to detect slices demonstrating infection, we noticed that in some cases the abnormalities are barely visible with the standard visualization setting (window center and window width). Those abnormalities have been detected by changing the image contrast (by adjusting the window center and width) manually by the radiologist. This limitation will arise the need to research on the optimal contrast and window level use in future studies. As another limitation, we can point to the retrospective study used in the data collection part of this research. Although the provided dataset is acquired with the utmost caution and inspection, a retrospective data collection might add inappropriate cases to the study at hand. The potential improvement to address this limitation could be the collaboration of more radiologists in analyzing and labeling the data to assess if the interobserver agreement is satisfying or not.

As a final note, unlike our previous work on the chest radiographs~\citep{Afshar2020a}, where we used a more imbalanced public dataset, the dataset used in this study contains a substantial number of COVID-19 confirmed cases making our results more reliable. Upon receiving more data from medical centers and collaborators, we will continue to further modify and validate the COVID-FACT by incorporating new datasets.


\bibliographystyle{frontiersinSCNS_ENG_HUMS}
\bibliography{references}

\end{document}